\documentclass[11pt]{article}
\usepackage{graphicx}
\usepackage{fancyhdr}
\RequirePackage{amsthm,amsmath,amsfonts,amssymb}
\RequirePackage[colorlinks,linkcolor=blue,citecolor=blue,urlcolor=black]{hyperref}
\RequirePackage{graphicx}
\usepackage[square, sort&compress, numbers]
{natbib}
\RequirePackage[ruled,vlined]{algorithm2e}
\usepackage{enumerate}
\usepackage[T1]{fontenc}

\usepackage{caption}
\RequirePackage[ruled,vlined]{algorithm2e}
\usepackage{enumerate}

\newcommand{\HH}[1]{}

\numberwithin{equation}{section}
\theoremstyle{plain}

\newlength\FHoffset
\fancyheadoffset{\FHoffset}
\title{\Large\bf  Neuronal Functional Connectivity Graph Estimation\\ with the {\sf R} package {\tt neurofuncon}}

\author{Lauren Miako Beede and Giuseppe Vinci\footnote{Corresponding Author E-mail: gvinci@nd.edu}\\{\footnotesize \it Department of Applied and Computational Mathematics and Statistics}\\ {\footnotesize \it University of Notre Dame, Notre Dame, Indiana, USA}}
\date{}

\usepackage{listings}
\usepackage{color}

\definecolor{palegreen4}{RGB}{84,139,84}
\definecolor{gray}{rgb}{0.5,0.5,0.5}
\definecolor{grey95}{RGB}{245,245,245}
\definecolor{firebrick3}{RGB}{205,38,38}

\begin{document}

\maketitle

\begin{abstract}\small 
Researchers continue exploring neurons' intricate patterns of activity in the cerebral visual cortex in response to visual stimuli. The way neurons communicate and optimize their interactions with each other under different experimental conditions remains a topic of active investigation. Probabilistic Graphical Models are invaluable tools in neuroscience research, as they let us identify the functional connections, or conditional statistical dependencies, between neurons. Graphical models represent these connections as a graph, where nodes represent neurons and edges indicate the presence of functional connections between them. We developed the {\sf R} package {\tt neurofuncon} for the computation and visualization of functional connectivity graphs from large-scale data based on the Graphical lasso. We illustrate the use of this package with publicly available two-photon calcium microscopy imaging data from approximately 10000 neurons in a 1mm cubic section of a mouse visual cortex.
\end{abstract}


\section{Introduction}

Brain connectivities cover a broad scope of both anatomical and functional connectivity. Anatomical connectivity is the network of neurons' physiological, synaptic connections, whereas functional connectivity is a network of neurons' statistical dependencies regardless of direct structural links.  Understanding functional connectivity is important for advances in the biomedical field for neuronal diseases, such as Parkinson's and Alzheimer's, and devices using brain-computer interfaces for prosthetics. Probabilistic graphical models let us describe the dependence structure of several random variables by means of a graph, where nodes represent the random variables and edges connect pairs of nodes to describe the dependence structure of the random variables \citep{lauritzen1996graphical,vinci2018adjusted,vinci2018adjustedcovariance,yuan2007model, friedman2008sparse}. In the context of neuronal functional connectivity, nodes represent neurons and edges represent functional connections. A useful framework in neuroscience is the Gaussian graphical model \citep{lauritzen1996graphical,vinci2018adjusted,vinci2018adjustedcovariance,yuan2007model, friedman2008sparse}, where the inverse $\Theta$ of the covariance matrix $\Sigma$ of neurons' activities encodes the conditional dependence for all pairs of neurons: $\Theta_{ij}=0$ if and only if the $i$-th and the $j$-th variables are conditionally independent (no edge connects them).  
Because of the large number of neurons, it is convenient to estimate $\Theta$ by using the regularized graph estimation approach Graphical Lasso (GLASSO) \citep{yuan2007model,friedman2008sparse}, which is the solution of the following optimization problem
\begin{equation}
    \hat\Theta(\lambda) ~=~ \arg\max_{\Theta\succ 0} ~~\log\det\Theta-{\rm trace}(\hat\Sigma \Theta)-\lambda\Vert \Theta\Vert_{1,\rm off}
\end{equation}
where $\hat\Sigma$ is the sample covariance matrix, $\Vert \Theta\Vert_{1,\rm off}=\sum_{i\neq j}|\Theta_{ij}| $, and $\lambda\ge 0$ is a tuning parameter that induces sparsity in the precision matrix estimate $\hat\Theta(\lambda)$. A larger $\lambda$ produces a sparser solution, hence an estimated graph with a smaller number of edges.
 
In this paper we present the {\tt neurofuncon} {\sf R} package, which implements functional connectivity graph estimation and visualization based on GLASSO. 
In section \ref{sec:neurofuncon}, we describe the {\sf R} package {\tt neurofuncon} and its functions in examples using publicly available calcium imaging data from \cite{stringer2019spontaneous}.  
Finally, in Section~\ref{sec:conclusion} we discuss future directions.

\section{The {\sf R} package {\tt neurofuncon}}
\label{sec:neurofuncon}

The {\sf R} package {\tt neurofuncon} can be used to estimate and visualize functional neuronal connectivity graphs from large-scale electrophysiological data, such as calcium imaging recordings. This package is available at \href{https://github.com/lbeede/neurofuncon}{https://github.com/lbeede/neurofuncon} and, after installation, it can be loaded with the following command:
\begin{lstlisting}
    library(neurofuncon)
\end{lstlisting}
We illustrate the use of {\tt neurofuncon} functions with a publicly available calcium imaging data set ``natimg2800\_M170604\_MP031\_2017-06-28.mat'' from \cite{stringer2019spontaneous}, which is available at \href{https://figshare.com/articles/Recordings_of_ten_thousand_neurons_in_visual_cortex_in_response_to_2_800_natural_images/6845348}{\url{https://figshare.com/articles/Recordings_of_ten_thousand_neurons_in_visual_cortex_in_response_to_2_800_natural_images/6845348}}.

\subsection{Dataset}
The file ``natimg2800\_M170604\_MP031\_2017-06-28.mat'' should be put in your working directory of R. Then, run the following code to load and prepare the data:

\begin{lstlisting} % [numbers=left]

    DATA <- R.matlab::readMat('natimg2800_M170604_MP031_2017-06-28.mat', drop=FALSE)
    
    TRACES <- DATA$stim[2][1][[1]][[1]]            
    POSITIONS <- DATA$med                          
    STIMULI <- as.integer(DATA$stim[3][[1]][[1]])  
    
    LAYER.IDS <- unique(POSITIONS[,3])  # finding all layers of the data
    NLAYER <- LAYER.IDS[1]  # first layer
    NLAYERS <- LAYER.IDS[1:7]  # first seven layers
    GGM.STIMULI.IDS <- c(184:316)   # a set of stimuli according to Stringer et al.
\end{lstlisting}

In this dataset, there are $n = 5880$ sample recordings from $d = 10079$ neurons in mouse V1 cortex. The object {\tt TRACES} is a $n \times d $ matrix of neuron activations in response to the 5880 stimuli contained in {\tt STIMULI}, where the rows represent the samples and the columns represent the neurons. The object {\tt POSITIONS} is a $ d\times 3$ matrix containing the 3D coordinates $(x, y, z)$ of the 10079 neurons.  
With these data now loaded, we can extract the neural responses from calcium imaging data points to visualize the neuronal architecture of the dataset. Figure~\ref{fig:plot3d} shows the three-dimensional positions of the neurons in the layers, and can be produced with the following code:

\begin{figure}
    \centering
    \includegraphics[width=0.3\textwidth, trim = 0 1.2cm 0 0]{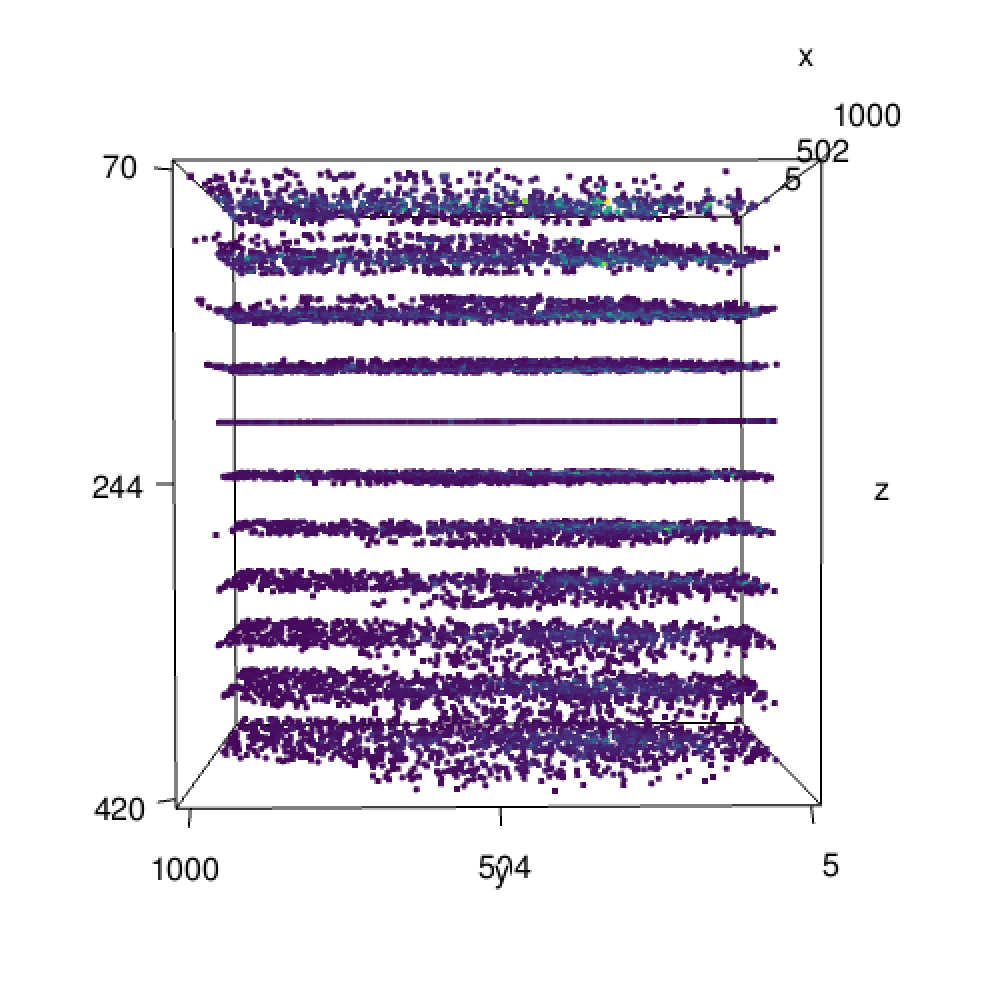}
    \includegraphics[width=0.3\textwidth, trim = 0 0 0 0]{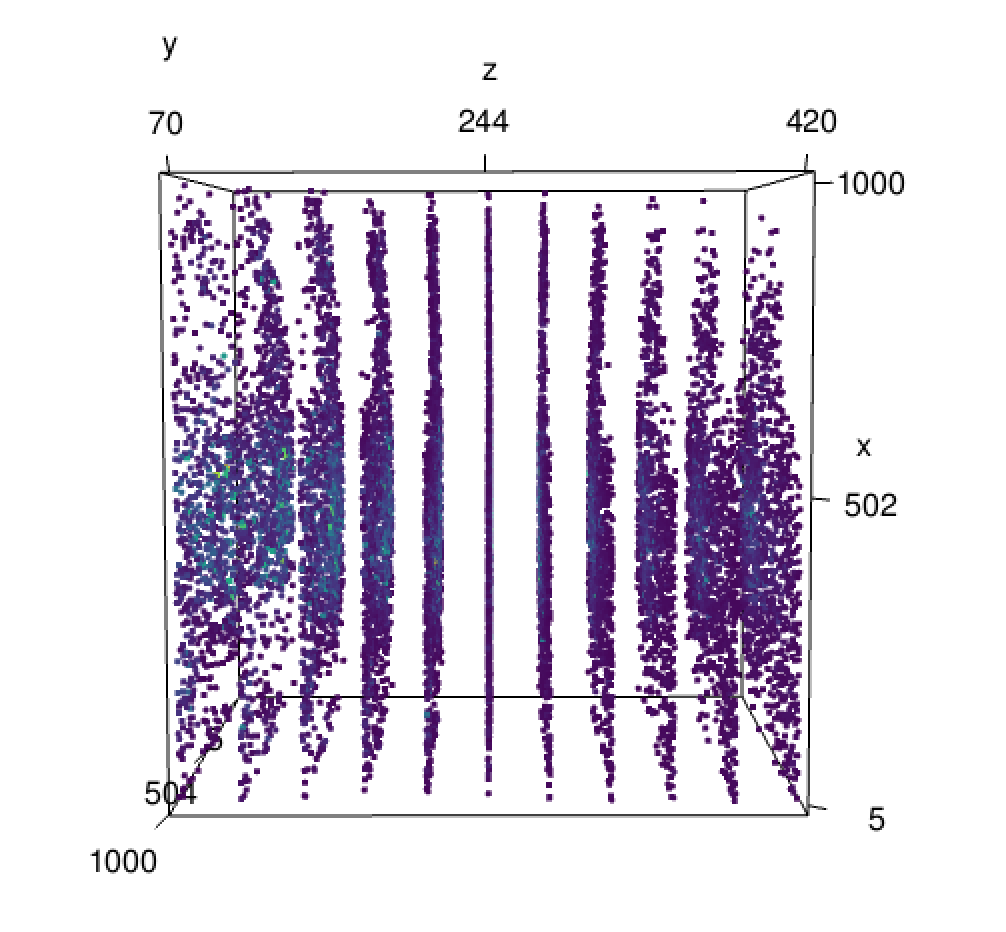}
    \includegraphics[width=0.275\textwidth, trim = 0 0 0 0]{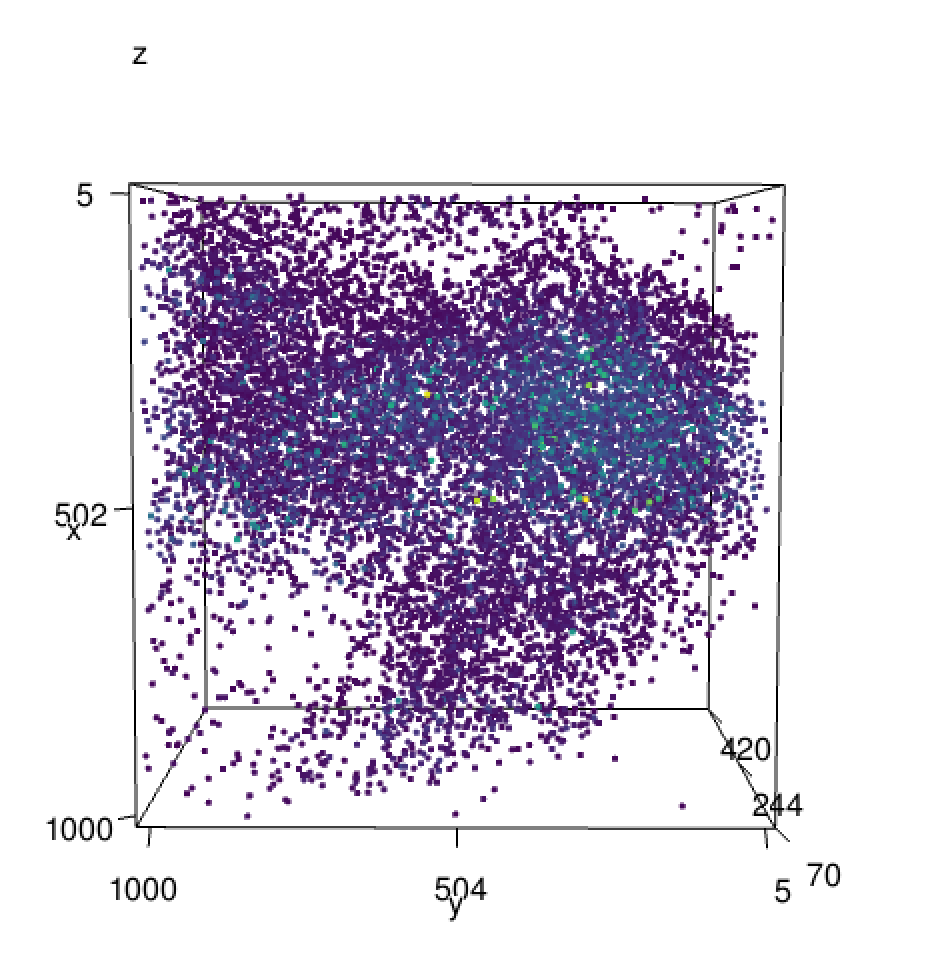}
    \caption{3D plots of average neuron activation in the 1mm cube across the 11 planes using data from \cite{stringer2019spontaneous}. Purple indicates low activation while yellow denotes high activation. 
    }
    \label{fig:plot3d}
\end{figure}

\begin{lstlisting}
    library(rgl)
    avg <- colMeans(TRACES)
    layers.avg <- data.frame(cbind(POSITIONS, avg))
    plot3d(layers.avg[,1], layers.avg[,2], layers.avg[,3],
       xlab = "x", ylab = "y", zlab = "z", 
       col = colourvalues::color_values(avg),
       axes = FALSE)
    box3d()
    axis3d(edge='x--',nticks=3, at = c(min(layers.avg[,1])+1, round(mean(c(min(layers.avg[,1])+1, max(layers.avg[,1])-4))), max(layers.avg[,1])-4))
    axis3d(edge='y--',nticks=3, at = c(min(layers.avg[,2])+1, round(mean(c(min(layers.avg[,2])+1, max(layers.avg[,2])-4))), max(layers.avg[,2])-8))
    axis3d(edge='z++',nticks=3, at = c(min(layers.avg[,3]),round(mean(c(min(layers.avg[,3])+1, max(layers.avg[,3])-4))),max(layers.avg[,3])))
    
\end{lstlisting}

In the following sections we describe the use of the functions {\tt neurofun2d} and {\tt neurofun3d} of the {\sf R} package {\tt neurofuncon}.

\subsection{2-dimensional neuronal functional connectivity graphs} 
 The function {\tt neurofun2d()} has five main arguments: {\tt traces}, a $n \times d $  matrix where $n$ is the number of samples and $d$ is the total number of neurons; {\tt position}, a $d\times3$ matrix containing the 3D coordinates $(x, y, z)$ of the $d$ neurons; {\tt all.stimuli.ids}, a vector containing the stimuli ids (numeric) related to the $n$ samples; {\tt nlayer}, the layer (z-axis) desired to create the GGM; and {\tt ggm.stimuli.ids}, the stimuli ids desired to create the GGM.

Additional arguments include: {\tt num.neurons}, the number of neurons with the highest activation means desired to create the GGM (default is {\tt 25}); {\tt num.edges}, the number of edges desired to create the GGM (default is {\tt 25}); {\tt rho.seq}, sequence of penalty parameters to use in the GLASSO selection (default is {\tt seq(5,0.001,-0.001)}); {\tt Plot}, a boolean value determining whether to plot the graph (default is {\tt TRUE}); and {\tt node.size}, the maximum size of the nodes in the graph (default is {\tt 4}).

\begin{figure}
    \centering
    \includegraphics[width=0.87\textwidth]{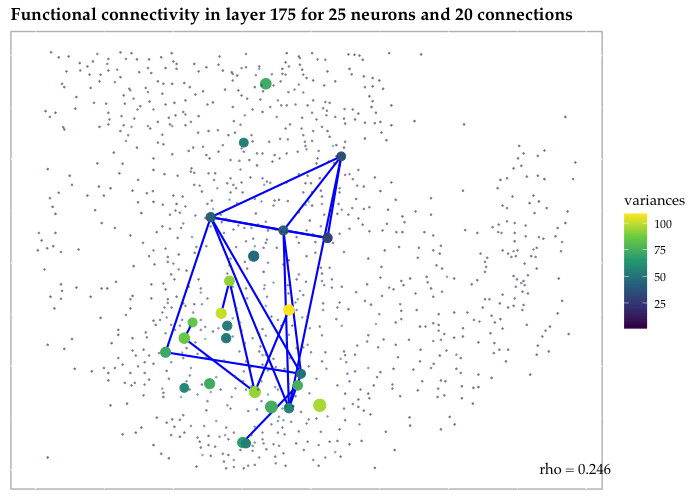}
    \caption{2D GGM graph of functional connectivity of the 25 neurons with the highest highest activation means and 20 connections in response to a set of stimuli.}
    \label{fig:neurofun2d_25x20}
\end{figure}

\begin{figure}
    \centering
    \includegraphics[width=0.87
\textwidth]{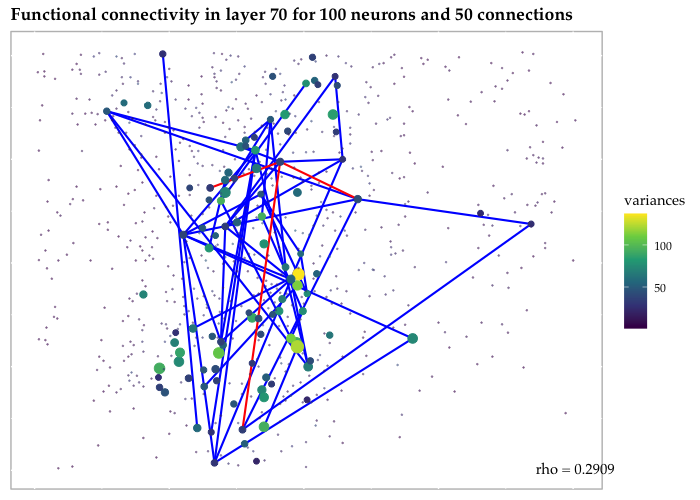}
    \caption{2D GGM graph of functional connectivity of the 100 neurons with the highest highest activation means and the 50 connections for mice in response to a set of stimuli.}
    \label{fig:neurofun2d_100x50}
\end{figure}

We now present examples of use of the function. Using the calcium imaging data we uploaded previously, the following code estimates and plots in Figure \ref{fig:neurofun2d_25x20} the conditional dependence graph with {\tt num.edges = 20} for the {\tt num.neurons = 25} neurons with the highest average activation:
\begin{lstlisting}
    neurofun2d(traces = TRACES, position = POSITIONS, all.stimuli.ids = STIMULI, nlayer = NLAYER, ggm.stimuli.ids = GGM.STIMULI.IDS, num.edges = 20)
\end{lstlisting}

The function {\tt neurofun2d()} returns a list containing the layer number, locations of the positive edges, locations of the negative edges, layer information, the optimal calculated rho value, the true number of edges, and the true number of neurons in addition to a plot. Use of the additional arguments and a larger number of {\tt num.neurons} and {\tt num.edges} can be implemented in a similar fashion as follows:  
\begin{lstlisting}
    neurofun2d(traces = TRACES, position = POSITIONS, all.stimuli.ids = STIMULI, nlayer = NLAYER, ggm.stimuli.ids = GGM.STIMULI.IDS, num.neurons = 100, num.edges = 50, rho.seq = seq(0.5, 0.0001, -0.0001), Plot = TRUE)
\end{lstlisting}
The code above produces Figure~\ref{fig:neurofun2d_100x50}. Run {\tt ?neurofun2d} for additional information and examples.

\subsection{3-dimensional neuronal functional connectivity graphs} 
The function {\tt neurofun3d()} has five main arguments similar to {\tt neurofun2d()}: {\tt traces}, a $n \times d $  matrix where $n$ is the number of samples and $d$ is the total number of neurons; {\tt position}, a $d\times3$ matrix containing the 3D coordinates $(x, y, z)$ of the $d$ neurons; {\tt all.stimuli.ids}, a vector containing the stimuli ids (numeric) related to the $n$ samples; {\tt nlayers}, a vector of the layers (z-axis) desired to create the GGM and {\tt ggm.stimuli.ids}, the stimuli ids desired to create the GGM.

Additional arguments include: {\tt num.neurons}, the number of neurons with the highest activation means desired to create the GGM (default is {\tt 25}); {\tt num.edges}, the number of edges desired to create the GGM (default is {\tt 25}); {\tt rho.seq}, sequence of penalty parameters to use in the GLASSO selection (default is {\tt seq(5,0.001,-0.001)}); {\tt Plot}, a boolean value determining whether to plot the graph (default is {\tt TRUE}); and {\tt node.size}, the maximum size of the nodes in the graph (default is {\tt 6}).

\begin{figure}
    \centering
    \includegraphics[width=0.365\textwidth, trim = 0 0 0 0]{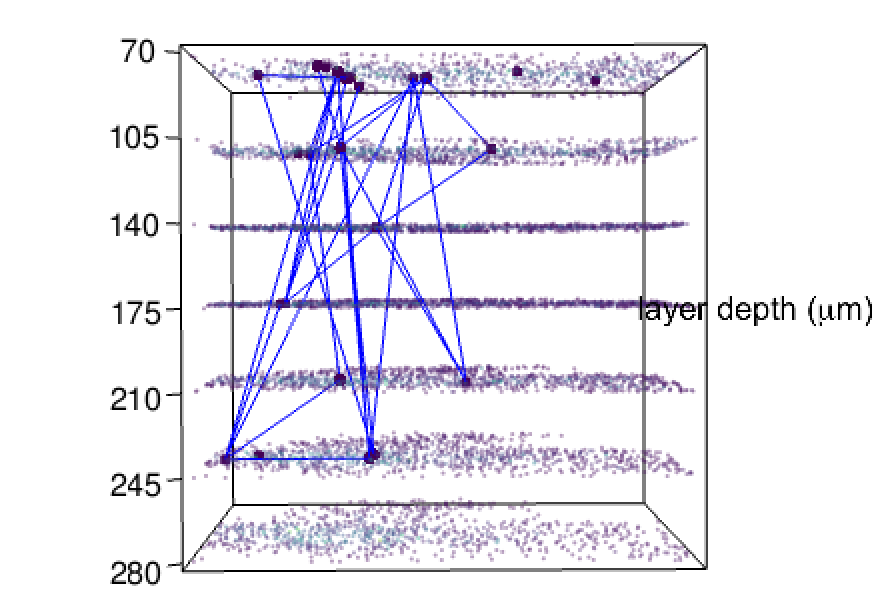}
    \hspace{0.1cm}
    \includegraphics[width=0.265\textwidth, trim = 0 1cm 0 0]{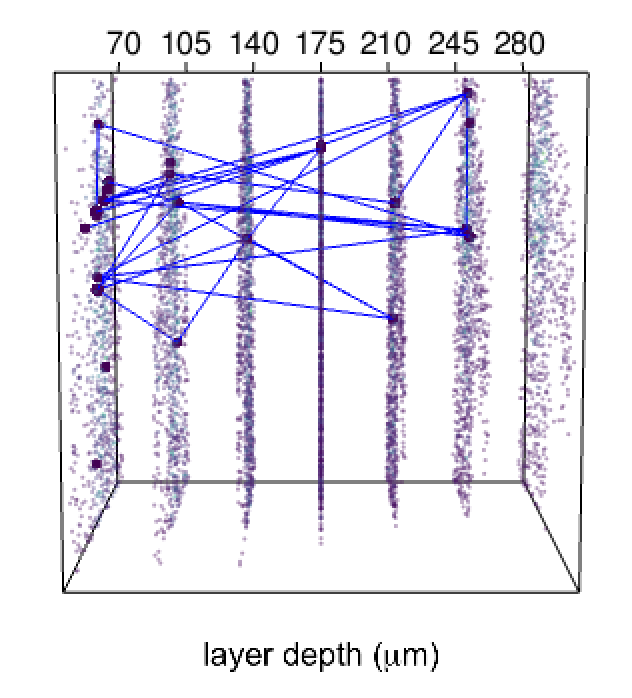}
    \includegraphics[width=0.34\textwidth, trim = -2cm -0.25cm 0 0]{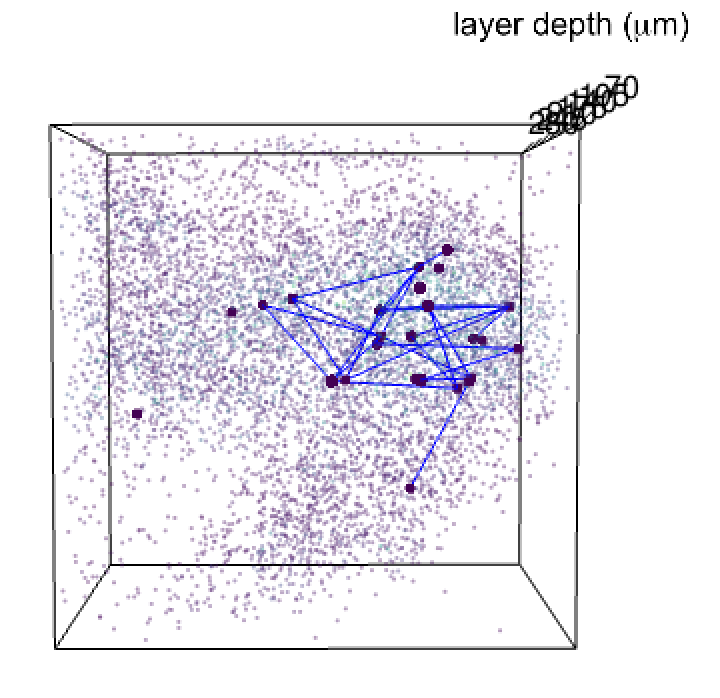}
    \caption{3D GGM graph of functional connectivity of the 25 neurons with the highest highest activation means and the 25 connections for mice in response to a set of stimuli.}
    \label{fig:neurofun3d_25x25}
\end{figure}

\begin{figure}
    \centering
    \includegraphics[width=0.355\textwidth, trim = 0 0 0 0]{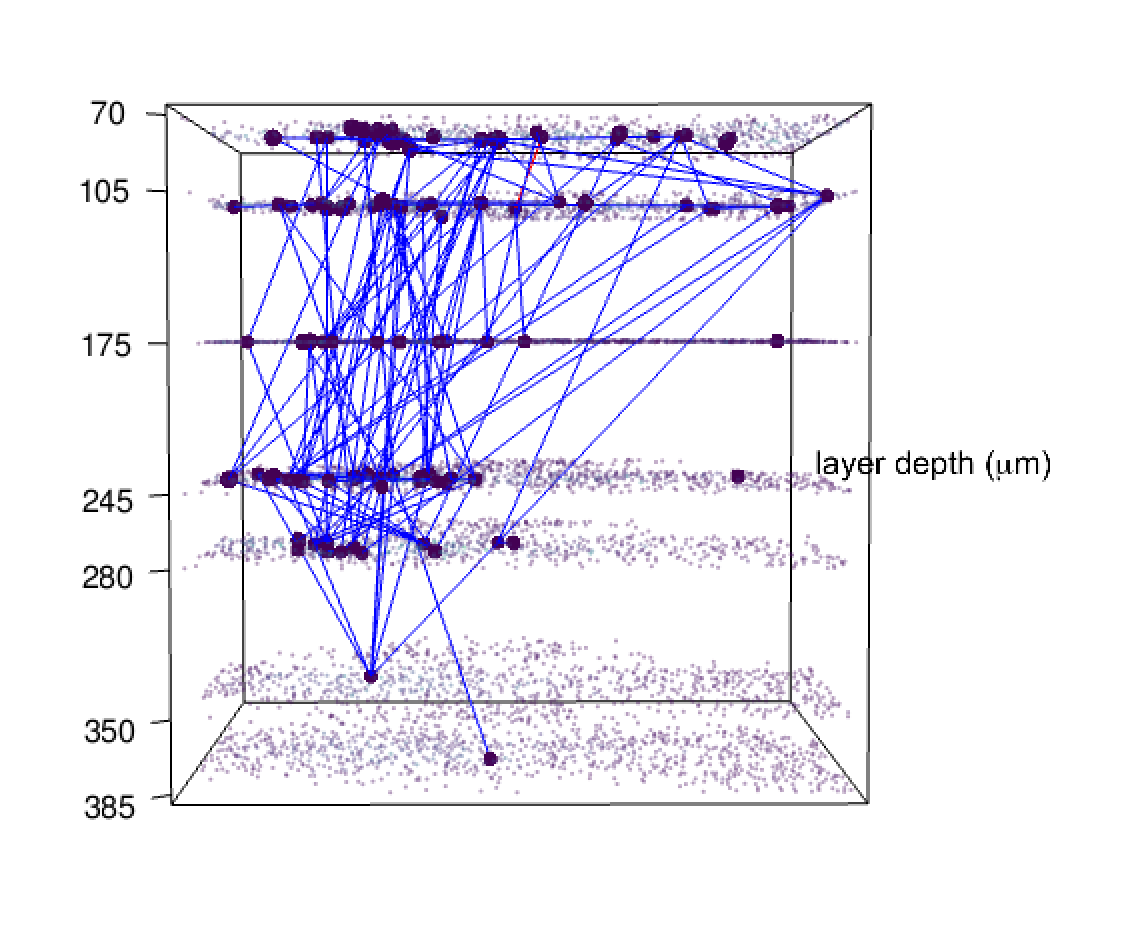}
    \includegraphics[width=0.3\textwidth, trim = 0 -0.6cm 0 0]{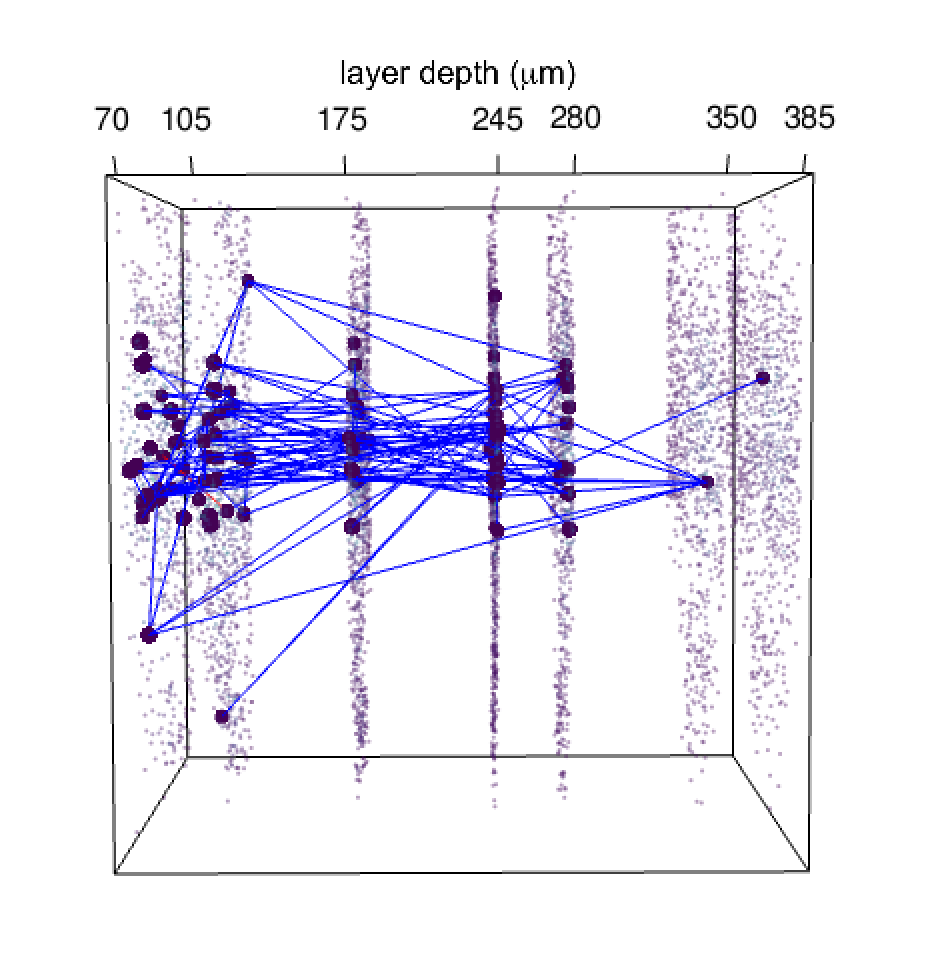}
    \includegraphics[width=0.325\textwidth]{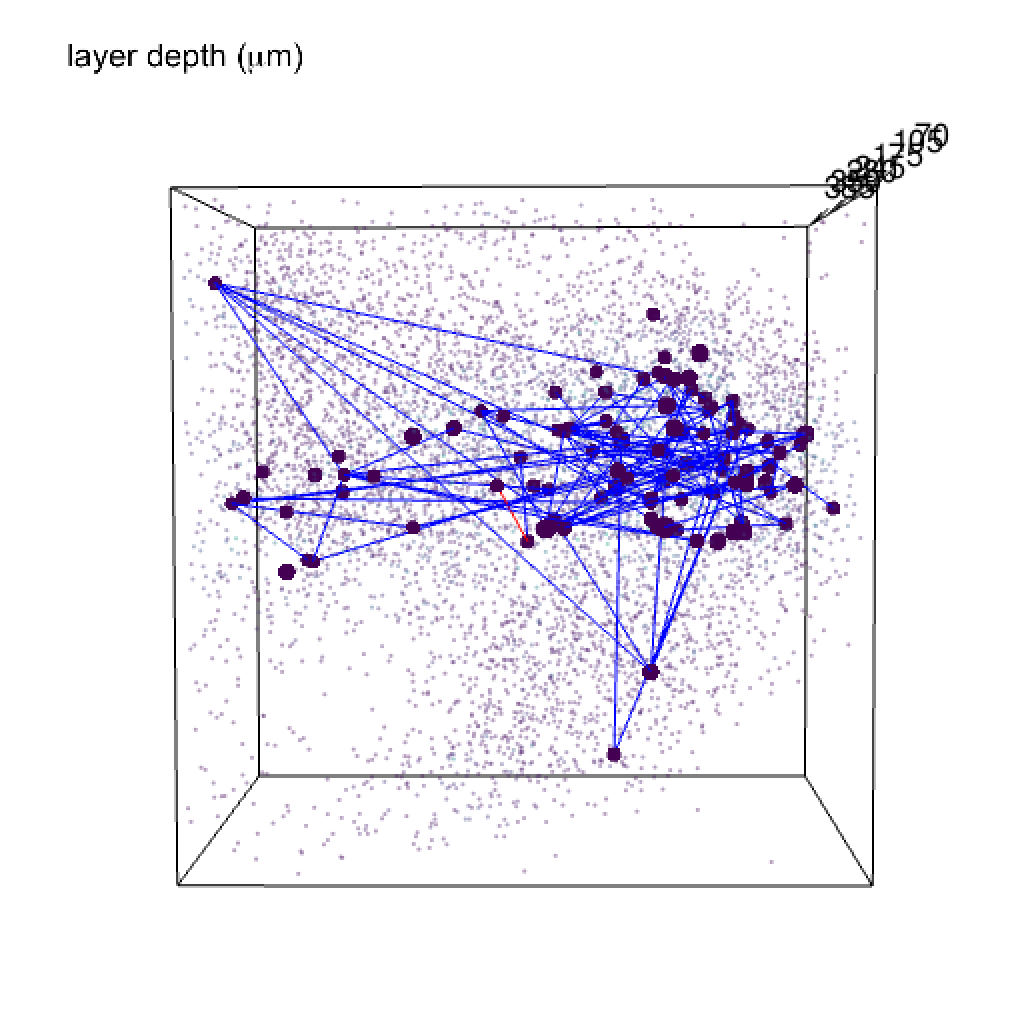}
    \caption{3D GGM graph of functional connectivity of the 100 neurons with the highest highest activation means and the 50 connections for mice in response to a set of stimuli. }
    \label{fig:neurofun3d_100x50}
\end{figure}

We now present examples of use of the function. The following code estimates and plots in Figure~\ref{fig:neurofun3d_25x25} the conditional dependence graph for the {\tt num.neurons = 25} neurons with the highest average activation in the 7 layers {\tt NLAYERS} we selected previously:
\begin{lstlisting}
    neurofun3d(traces = TRACES, position = POSITIONS, all.stimuli.ids = STIMULI, nlayers = NLAYERS, ggm.stimuli.ids = GGM.STIMULI.IDS, num.edges = 20)
\end{lstlisting}

The function {\tt neurofun3d()} returns a list containing the layer number, locations of the positive edges, locations of the negative edges, layer information, the optimal calculated rho value, the true number of edges, and the true number of neurons in addition to a plot. Use of the additional arguments and a larger number of {\tt num.neurons} and {\tt num.edges} can be implemented in a similar fashion as follows:  
\begin{lstlisting}
    neurofun3d(traces = TRACES, position = POSITIONS, all.stimuli.ids = STIMULI, nlayer = NLAYERS, ggm.stimuli.ids = GGM.STIMULI.IDS, num.neurons = 100, num.edges = 50, rho.seq = seq(0.25, 0.1, -0.0001), Plot = TRUE, node.size=10)
\end{lstlisting}
The output is shown in Figure \ref{fig:neurofun3d_100x50}. Run {\tt ?neurofun3d} for additional information and examples.

\section{Conclusion}\label{sec:conclusion}

We presented the {\sf R} package {\tt neurofuncon} for the estimation and visualization of functional neuronal connectivity graphs from large-scale neuronal data sets. We illustrated the use of the {\tt neurofuncon} functions by analysing publicly available calcium imaging data \citep{stringer2019spontaneous} recorded from about ten thousand neurons in mouse visual cortex. We expect the {\sf R} package {\tt neurofuncon} to allow neuroscientists and practitioners to visualize functional connectivity graphs efficiently. We will keep updating and improving the {\tt neurofuncon} package regularly on our GitHub webpage, by adding new functions for computation and visualization of functional connectivity graphs.

\bibliographystyle{unsrt}
\bibliography{references}

\end{document}